**Tempering and few-shot calibration of pathology foundation models improve digital biomarker reproducibility across slide scanners**

Wen, Frank T[1*]; Rosen, Kyler[2*]; Ramesh, Siddhi[2]; Koo, Michael[2]; Kochanny, Sara[1]; Jain-Liu, Sasha[1]; Dyer, Emma[1,3,4]; Pearson, Alexander T[1,5†], Dolezal, James M[2,6*†]

[1]The University of Chicago, Chicago, IL
[2]Slideflow Labs, Chicago, IL
[3]The Dana Farber Cancer Institute, Boston, MA
[4]Harvard T.H. Chan School of Public Health, Boston, MA
[5]Chan Zuckerberg Biohub Chicago LLC, Chicago, IL
[6]Geisinger Cancer Institute, Danville, PA
[*]These authors contributed equally
[†]Correspondence to: jdolezal@geisinger.edu; apearson5@medicine.bsd.uchicago.edu

## Abstract

Pathology foundation models underpin a growing number of histology-based cancer biomarkers, but sensitivity to slide scanner-specific image signatures degrades cross-center reliability and reproducibility. We developed a foundation model fine-tuning method that improved consistency in head-and-neck cancer human papillomavirus (HPV) prediction, colorectal cancer microsatellite instability (MSI) identification, prostate cancer grading and breast cancer phenotyping. This tempering method decreased cross-scanner prediction differences by 67.5% for HPV (95% CI 59.0–73.9%), 69.3% for MSI (95% CI 51.9–76.5%) and 63.4%–81.2% for breast signatures. Cross-scanner prostate grade agreement rose from 42.5% to 92.7%. Under full scanner–HPV confounding during training, AUROC fell from 0.80 to 0.68 with original model features but only from 0.86 to 0.84 with tempering. Unlike post hoc feature calibration, tempering used only public data yet matched or exceeded on-site calibration. These findings support tempering to reduce scanner dependence of pathology foundation models.

## 1 Introduction

Deep learning applied to routine hematoxylin and eosin (H&E) slides has produced digital cancer biomarkers that infer diagnosis, subtype, grade, molecular alterations, recurrence risk and treatment benefit[1–7], and several are now authorized as medical devices for use with specified slide scanners[8–10]. Pathology foundation models, pretrained on millions of tissue images, have become the common starting point for such biomarkers[11–13], with downstream models trained to aggregate foundation model features into a slide- or patient-level output[14,15]. Although foundation models serve as powerful starting points for identifying relevant biological features, they often also encode scanner-specific signal unrelated to tissue biology, reflecting features related to the method of image capture such as scanner optics, spatial sampling, color rendering, sharpness and compression. These encoded nonbiological features pose latent risks for bias, confounding, and model drift when biomarkers are deployed on diverse scanner hardware.

Recent rescanning studies show that foundation-model features and downstream predictions often change when only the scanner changes[16,17], extending earlier evidence of scanner-induced domain shift in task-specific networks[18,19]. Critically, scanner-dependent prediction shifts may not be detected by common discrimination-based metrics (i.e. AUROC) and only manifest when a fixed decision threshold is applied, which better represents a practical deployment scenario[16]. Confounding during training also can introduce bias: when acquisition site or scanner correlates with outcome, models may rely partly on image acquisition signatures to make predictions, rather than on the intended biology[20–22].

Proposed remedies intervene across different points in the modeling process, and include stain normalization and augmentation of the input[23,24], correction of pretrained features[25], robustness-aware task-specific training[17,26], and fine-tuning of the foundation model itself[27]. Post hoc correction of pretrained features through calibrated feature-space transforms can reduce scanner batch effects but requires pixel-registered rescans of the same tissue on every deployment scanner, which are technically and logistically challenging to acquire[25]. Improving the foundation model is attractive because every downstream biomarker would inherit the correction. Whether encoder fine-tuning and post hoc correction are substitutes or complements, and whether encoder-level robustness improves the reproducibility of clinical predictions and protects against acquisition shortcuts, has not been tested across cancer types.

Here we introduce "tempering", a fine-tuning method for improving foundation-model consistency using spatially registered images of the same tissue acquired on different scanners, applied to the H-optimus-0 foundation model[28]. We report the resulting tempered encoder, Mettle-1.5, and compare it with few-shot "calibration", which distills cross-scanner corrections learned from a small set of paired specimens into a metadata-free head, alone and in combination. Using paired rescans and a controlled confounding experiment, we evaluated all four combinations across four clinical applications: HPV status in head-and-neck cancer, microsatellite instability (MSI) in colorectal cancer, gene-expression signatures in breast cancer and Gleason grading in prostate cancer. We find that correcting scanner dependence in the shared encoder, using only public data, makes predictions reproducible across scanners in all four applications, matching or exceeding calibration fitted on the evaluation scanners, and protects downstream training against scanner shortcuts while preserving biological performance.

## 2 Results

### *2.1 Tempering and calibration study design*

Tempering applies self-supervised fine tuning to a foundation model so that registered views of the same tissue from different scanners share one representation, anchored to the original model's embedding of a reference view (**Fig. 1a**). Low-rank adapters[29] and a content-conditioned correction head are trained jointly on real cross-scanner tissue pairs and on synthetic views that simulate acquisition differences. We tempered H-optimus-0 on public registered collections, SCORPION[30], PLISM[31] and ten prostate specimens from the PAR dataset[32]. We name the resulting encoder Mettle-1.5 (**Fig. 1b**).

Few-shot calibration is a post hoc step that applies a learned correction between scanners (**Fig. 1c**). Using registered rescans across paired scanners, we distilled corrections learned from a small calibration set into one content-conditioned head. This yields a model that is calibrated to scanners of interest, rather than applying a fixed transform for each scanner[25]. The calibration set comprised the ten PAR tempering specimens and 15 University of Chicago cross-scanner specimens from the evaluation scanners and sites. Fitting the head to each encoder yields a 2×2 design with four arms: stock (unmodified H-optimus-0), calibrated Stock, Mettle-1.5, and calibrated Mettle-1.5 (**Fig. 1d**).

Tempering and calibration cases were excluded from all evaluations (**Fig. 1c**). Mascaret, a contemporaneous robustness-tuned foundation model[27], served as a comparator.

### *2.2 Tempering increases the consistency of predictions for a given slide scanned across different scanners for HPV status in head and neck cancers*

We first tested whether tempering makes downstream predictions for the same slide more consistent across scanners, using HPV status in head and neck squamous cell carcinoma. MIL models of HPV status were trained on 345 head-and-neck slides from The Cancer Genome Atlas (TCGA)[33] with each arm's features and evaluated on 88 University of Chicago head-and-neck squamous cell carcinoma specimens (46 HPV-positive) scanned on a Leica Aperio GT450 (reference), Leica Aperio AT2, Grundium Ocus 20, and Olympus VS200 scanners (239 total reference–rescan pairs). We measured discrimination by the area under the receiver operating characteristic curve (AUROC), and consistency by scanner discrepancy (defined as the root-mean-square difference between rescan and reference probabilities) and by calls changed at a threshold of 0.5. Consistency metrics were computed using the GT450 as the reference scanner.

Downstream MIL models trained using stock features had stable discrimination across scanners but the same specimens received different scores and calls. Among 63 specimens with all four scans, AUROC ranged from 0.807–0.812 (**Fig. 2a**), yet scores shifted between scanners (**Fig. 2b**). Sensitivity at a threshold of 0.5 ranged from 56% on the Grundium to 76% on the AT2, and specificity from 69% to 86% (**Fig. 2a**). Overall, 15 of 239 rescans (6.3%) changed HPV prediction (**Fig. 2b,e**).

Training on tempered features reduced scanner discrepancy by 68% (95% confidence interval [CI] 59–74%; $P < 0.001$), from 0.117 with stock to 0.038 with Mettle-1.5 (Fig. 2c). AUROC (0.799–0.803), sensitivity (79–82%) and specificity (66–72%) were stable across scanners (**Fig. 2a**). Three of 239 rescans (1.3%) changed call (**Fig. 2e**), and the reduction in changed calls held across nearly all thresholds (**Fig. 2f**). Discrimination was also preserved in 362 further University of Chicago cases (160 HPV-positive) scanned on an Aperio XT device (AUROC 0.867 with stock, 0.864 with Mettle-1.5; difference −0.003, 95% CI −0.015 to 0.008).

Calibration reduced discrepancy less than tempering did. Calibrated Stock, fitted on acquisitions from the same site and scanners, reached 0.063 (difference from Mettle-1.5 −0.025, −0.037 to −0.013; $P < 0.001$), its sensitivity still varied across scanners (65–76%; $P = 0.029$), and 10 rescans changed call ($P = 0.040$ versus uncalibrated Mettle-1.5). Calibrating Mettle-1.5 lowered discrepancy further, to 0.022 ($P < 0.001$), with 4 rescans changing call (**Fig. 2c,e**).

### *2.3 Tempered features resist scanner–disease confounding during downstream classifier training*

We next tested whether scanner signatures also distort learning when scanner and outcome are systematically confounded using the same cohort of slides and scanners. Within three scanner pairs of the paired cohort (AT2 with Grundium, Leica, or Olympus), we enforced the scanner–HPV association (Cramér's V) in the MIL training cohort from none (0) to complete (1) with cases and labels fixed (**Fig. 3a**). Each arm's 450 models were evaluated on 436 TCGA patients (50 HPV-positive).

Models trained on tempered features lost less discrimination when the training data were confounded than models trained on stock features. Stock-feature models lost 0.125 AUROC (95% CI 0.084–0.165) between no confounding and complete confounding (Cramér's V of 0 and 1), whereas Mettle-1.5 and calibrated Stock lost 0.020 and 0.015 (difference 0.006, −0.016 to 0.029; $P = 0.64$) and calibrated Mettle-1.5 lost 0.004. Mettle-1.5's AUROC was higher than calibrated Stock's both without and with complete

confounding (0.857 versus 0.811, P = 0.017; 0.040 higher under complete confounding, 0.009–0.071, P = 0.010; **Fig. 3b**), whereas calibration of stock alone did not improve AUROC without confounding (0.811 versus 0.801 for stock; P = 0.15).

Tempering also substantially reduced a scanner-driven shift in how many patients the models called HPV-positive under complete confounding, an effect that AUROC, as a measure of ranking, does not capture (**Fig. 3c**). Stock-feature models called 79% of TCGA patients HPV-positive when all HPV-positive training cases had been imaged on the AT2, but 11% when they had been imaged on the partner scanner (prevalence 11.5%), a 68-percentage-point difference that fell to 23 points with calibrated Stock and to 0.9 points (95% CI −2.1 to 4.2) with Mettle-1.5 (P < 0.001 versus calibrated Stock).

Mascaret[27], run through the same design, also limited the AUROC loss (0.036), but its calls and mean predicted probabilities still differed between the direction of confounding (by 40 percentage points and 0.10). Its AUROC was below that of Mettle-1.5 both under complete confounding (by 0.054, 95% CI 0.015–0.094; P = 0.005) and without confounding (by 0.038). Thus, tempering distinguishes from calibration and Mascaret by its superior resistance to changes in the call rate under scanner confounding.

### *2.4 Tempering makes prostate grades reproducible across scanners*

To test prostate grading, we trained attention-based MIL models to predict prostate cancer grade on 568 TCGA[34] and SICAPv2[35] slides and applied them to prostate biopsies from the PAR dataset[32] scanned on a Leica Aperio GT450, a Grundium Ocus 40, and a Hamamatsu NanoZoomer 2.0 HT. Grading assessments from up to three different pathologists were available. After excluding development specimens and linked cases, 159 specimens from 100 case families remained.

Compared with stock, MIL models trained using Mettle-1.5 features increased cross-scanner quadratic weighted kappa from 0.72 to 0.95 (difference 0.23, 95% CI 0.19–0.28; P < 0.001) and exact grade agreement between scanners from 42.5% to 92.7% (**Fig. 4a,b**). Every Mettle-1.5 model exceeded every stock model (**Fig. 4b**). Calibrating Mettle-1.5 added little (93.8% and 0.95; P = 0.69). Mettle-1.5 also exceeded calibration alone (80.2% and 0.87; kappa difference 0.080, 0.047–0.120; P < 0.001).

Tempering also stabilized clinical grade-group calls (**Fig. 4c**). At thresholds of grade group ≥2, ≥3, ≥4 and 5, stock calls changed with the scanner in 10–24% of specimens, compared with 1.3–3.4% with Mettle-1.5 and 4.4–8.7% with calibrated Stock (Mettle-1.5 vs calibrated Stock, P ≤ 0.012). Stock disagreement was systematically biased, assigning grade group 5 to 12% of Grundium scans compared to 34% of Hamamatsu scans of the same specimens, whereas Mettle-1.5 did so on 42% and 44% respectively (**Fig. 4a**).

Weighted kappa against each pathologist also improved: against pathologist 1 from 0.614 with stock to 0.688 with Mettle-1.5 (difference 0.075, 0.022–0.131; P = 0.004), and against pathologists 2 and 3 by 0.074 (P = 0.007; 157 specimens) and 0.133 (P < 0.001; 39 specimens; **Fig. 4d**). Inter-pathologist agreement (weighted kappa 0.48-0.78) was below Mettle-1.5's cross-scanner agreement (kappa 0.95) and within the range of stock (0.72), indicating that models trained on tempered features agreed across scanners more often than pathologists agreed with each other. Mettle-1.5 assigned grade group 5 to 42-44% of scans, more often than pathologist 1 (24%). The disagreement with pathologist 1 was concentrated in specimens that pathologist 1 graded 4, most of which were graded 5 by the other two pathologists (15 of 27 by pathologist 2 and 6 of 7 by pathologist 3; **Fig. 4e**).

A tempered control trained without PAR data (but with rescans of 18 University of Chicago specimens) retained 77% of the cross-scanner gain (kappa 0.90 versus 0.87 with calibration fitted on PAR rescans, P =

0.12; exact agreement 85.7% versus 80.2%, P = 0.005) and upgraded grade group 4 specimens as often as Mettle-1.5, but did not improve agreement with any pathologist (pathologist 1: 0.64 versus 0.61; P = 0.43; **Fig. 4b,d,e**).

### *2.5 Consistency of downstream models trained on tempered features extends to colorectal and breast biomarkers*

We next extended our analysis to colorectal cancer, training models to predict microsatellite instability (MSI) on 473 TCGA slides from 25 sites,[36] testing on whole slides of 62 University of Chicago specimens (13 MSI-high) with the reference and all three rescans (186 pairs; **Fig. 5a**).

Compared with stock, Mettle-1.5 reduced MSI scanner discrepancy from 0.067 to 0.021 (69%, 95% CI 52–76%; P < 0.001; **Fig. 5b**); calibration alone reduced it to 0.044, and calibrating Mettle-1.5 to 0.011. Calls at a threshold of 0.5 changed on six of 186 rescans with stock, two with calibrated Stock, one with Mettle-1.5 and none with calibrated Mettle-1.5. Mettle-1.5 and calibrated Stock did not differ significantly in call changes (difference −0.023, 95% CI −0.048 to 0.003; P = 0.22): calibration kept 50 of 62 specimens more stable, whereas tempering limited the largest specimen discrepancy (**Fig. 5c**). Mean AUROC across acquisitions was 0.972 with stock and 0.968 with Mettle-1.5 (**Fig. 5b**).

For breast cancer, we tested whether a downstream model trained on stock features becomes more consistent across scanners when applied to tempered features without retraining. We assessed two published breast cancer models trained to predict gene-expression signatures from H&E[5]: a bronchioid signature[37] that predicts recurrence and chemotherapy benefit in early breast cancer[5], and an approximation of the 21-gene recurrence score (Oncotype DX)[38]. Both were trained on stock H-optimus-0 features and their training data were not available, so we deployed frozen models using using each arm's features as inputs and measured the consistency of model outputs across scanners at 100 aligned tissue locations per acquisition in 77 specimens (**Fig. 5a**). Because the frozen models were not fitted to tempered or calibrated features, their output scale differed between arms. Discrepancy was therefore normalized across arms by dividing by the standard deviation of scores from acquisitions on the reference scanner for each arm.

Scores from tempered features agreed with those from stock features on the reference acquisition (Lin's concordance 0.92 bronchioid, 0.86 recurrence score; **Fig. 5d**). Mettle-1.5 reduced scaled breast cross-scanner discrepancy relative to stock by 63% (95% CI 56–69%) for the bronchioid signature and 81% (79–84%) for the recurrence score, more than calibrated Stock (43% and 67%; P < 0.001). Calibrated Mettle-1.5 reduced discrepancy by 79% and 89% (**Fig. 5a**). With stock features, scores shifted systematically on every rescan device. Residual discrepancies using tempered features were concentrated in Grundium acquisitions (−0.19 and +0.23 reference s.d.), which calibration removed (**Fig. 5e**).

### *2.6 Tempering reduces discrepancy under perturbations and on scanners absent from calibration*

Tempering reduced discrepancy under synthetic image perturbations; calibration did not. On 50 TCGA slides perturbed by JPEG re-encoding, Macenko stain transfer and fixed color transforms, HPV discrepancy between each perturbed image and its unperturbed original fell by 80–81% with Mettle-1.5 after re-encoding and by 67% across stain and color changes, whereas calibration changed it by −25% to +19%. Calibration corrects the scanner offsets it was fitted on, not image degradation in general.

Mettle-1.5 was also tested on a small auxiliary set of six head-and-neck specimens from University of Chicago scanned on three scanners absent from calibration (an Aperio XT, a 3DHISTECH Pannoramic SCAN 150, and a Philips IntelliSite UFS). MIL models trained on tempered features were more consistent than those trained on Stock, with mean discrepancy 0.037 with stock, 0.031 with calibrated Stock, 0.020 with Mettle-1.5 and 0.015 with calibrated Mettle-1.5.

### *2.7 Mettle is more robust on public benchmarks and preserves biological performance*

On PathoROB[21] (a benchmark used during development), the robustness index—which measures whether embeddings group by biological class rather than by medical center—increased from 0.801 with stock to 0.934 with Mettle-1.5. On the CRoMa distributional-robustness benchmark[39], Mettle-1.5 ranked second of 26 foundation models, after Mascaret[27], with the best lower-tail margin, while stock ranked 15$^{th}$. Mettle-1.5 matched or exceeded stock on HEST spatial transcriptomics[40], BACH breast histology[41], and three EVA tasks[42].

## 3 Discussion

Pathology foundation models reduce complex image inputs to high-dimensional embeddings for use in diverse downstream tasks. However, foundation model embeddings still encode acquisition signatures that downstream models may confound with biological signal. Here, we show that tempering a pathology foundation model by low-rank adaptation using public registered cross-scanner images attenuates the confounding effects of acquisition method. Across multiple tumor types and clinical endpoints, prediction scores for the same tissue scanned on different scanners were more similar using tempered embeddings than stock embeddings with no loss of discrimination. Additionally, downstream models trained using tempered embeddings were more resistant to scanner-outcome confounding during training. Tempered embeddings also gave more consistent predictions across scanners than stock embeddings that underwent few-shot calibration on registered multi-scanner acquisitions of evaluation-site slides.

Unlike existing approaches to mitigate acquisition confounding[17,23–25], tempering does not require acquisitions from the deployment site or robustness-specific training of each downstream model. MIL models trained on Mettle-1.5 features to predict head and neck cancer HPV status, colorectal cancer MSI status, and prostate cancer grade became more consistent across scanners compared to models trained on stock features despite having no robustness-specific adaptation during MIL training. Moreover, models trained using tempered features scored the same slide more consistently across scanners than models trained on calibrated stock features, even though calibration was fit to images from evaluation site and devices, while tempering used images from other sites and devices (albeit of the same manufacturers). For practical deployment, these results suggest that a digital biomarker built on a tempered encoder can be used across laboratories without a site-specific correction step.

Beyond facilitating consistency across scanners at inference, tempering reduces the tendency for downstream models to fit scanner-associated variation. Under systematically confounded training cohorts, models trained to detect HPV status in head and neck cancers using tempered features were less dependent on scanner than models trained using stock features. Calibration of stock features protected AUROC under confounding as well as tempering did but did not remove the dependence of calls on scanner. Tempering potentially influences what the downstream model learns by changing availability of scanner-associated features, as which features a model fits depends on their availability as well as their predictivity[43].

A useful property of tempering is that downstream models trained on features from a stock foundation model can be applied to features from a tempered version without modification. The breast signature models, trained on stock features and applied unchanged, produced the same signatures with greater consistency across scanners (even more than on calibrated stock features). However, model outputs shifted in scale, so a model moved to tempered features requires revalidation of its operating threshold. Overall, this property allows the developer of a biomarker built on an existing encoder to assess a tempered encoder on a biomarker's own validation cohort before adopting it, to determine whether cross-scanner consistency improves and whether the threshold need be adjusted.

Tempering also improves the metric of scanner dependence most relevant to clinical deployment, namely calls at a fixed threshold. A deployed biomarker applies a fixed threshold to test cases slides, so a scanner effect that shifts scores can change which slides are called positive even when the ordering is unchanged. Assessment of models by AUROC may fail to capture this effect[16]. In our experiments, tempered features improved fixed-threshold agreement between scanners across models trained to predict head and neck cancer HPV status, colon cancer MSI status, and prostate cancer grade, without change in AUROC. These results also reinforce that validation of biomarkers for use across sites should include paired acquisitions and fixed-threshold agreement.

This study has limitations. These results demonstrate reproducibility primarily across scanners whose manufacturers were represented in the tempering data and at the site where calibration was fitted. Transfer to other devices is less certain. No Olympus device was used in tempering, and on the VS200 tempering reduced discrepancy for MSI and the breast signatures but not for HPV. We tempered one foundation model, and model behavior may differ for encoders with other pretraining. All evaluations were retrospective, so whether these gains hold in prospective multicenter use remains to be shown.

## 4 Methods

### *4.1 Study design and ethics*

This retrospective study compared stock H-optimus-0 and its tempered version, Mettle-1.5, each with and without few-shot calibration (a 2×2 factorial design), using paired acquisitions, controlled training experiments and public benchmarks. The final checkpoint was selected using PathoROB. No clinical evaluation cohort, outcome or downstream prediction informed training or selection. This study was approved by the University of Chicago Institutional Review Board (protocol IRB 20-0238) and was conducted in accordance with the Declaration of Helsinki. Archival hematoxylin and eosin slides and associated clinical and molecular data from patients treated at the University of Chicago were analyzed retrospectively. The requirement for informed consent was waived because the research involved no more than minimal risk and used existing specimens and records.

### *4.2 Tempering*

Tempering adapts a pretrained foundation model so that images of the same tissue from different acquisitions yield matching representations while the biological structure encoded by the pretrained model is retained. Starting from H-optimus-0 (ViT-g/14, 1,536-dimensional class-token output)[28], we kept the pretrained weights frozen and trained low-rank adapters (LoRA)[29] on the attention and feed-forward projections of the transformer, with ranks in the low hundreds, together with a content-conditioned correction head.

*Training data.* Each batch of 96 tile pairs drew approximately one-third each from registered cross-scanner pairs in SCORPION (five scanners)[30], cross-stain and same-section pairs in PLISM[31] and registered cross-scanner pairs from ten PAR specimens[32]. In about one in five pairs, the comparison view was the reference tile itself. Comparison views were further perturbed with synthetic transformations drawn from four families of acquisition-inspired changes: color and stain rendering, sharpness and blur, resampling, and JPEG compression.

*Objective.* For two views of the same tissue location, a reference view $r$ and a comparison view $c$, the adapted encoder $f_\theta$ followed by the correction head $h$ embedded the comparison view, and a frozen copy of the original model $f_0$ embedded the clean reference view. The alignment loss was the L1 distance between class-token embeddings, $\mathscr{L} = \|h(f_\theta(c)) - f_0(r)\|_1$, so that the target remained anchored to the original representation. An auxiliary invariance term with a small, fixed weight aligned the embeddings of synthetic scanner-like views of the same tile. Training used AdamW with a peak learning rate of $10^{-4}$, linear warmup and cosine decay, for approximately one thousand optimizer steps, requiring approximately five GPU-hours on a single NVIDIA A100.

*Correction head.* The head adds a content-dependent correction to each embedding $z$: $h(z) = z + \Sigma_k\, \alpha_k(z)\, z\, \mathbf{U}_k\mathbf{V}_k^T + b(z)$, where $\mathbf{U}_k\mathbf{V}_k^T$ are low-rank matrices (tens of components, each of rank below ten) and the component weights $\alpha_k(z)$ and offset $b(z)$ are small networks of $z$. The head was initialized to the identity and uses no scanner or other metadata. Adapters and head were merged into a single model file operating on 224 × 224-pixel tiles.

*Checkpoint selection.* Recipe choices during development and the final checkpoint were compared on the mean PathoROB robustness index[21], with BACH balanced accuracy[41] as a regression check on biological signal; no clinical evaluation cohort, outcome or downstream prediction was used. In the final round, the untuned base model (robustness index 0.933) and 19 continuation checkpoints (0.921–0.934) were compared, and the selected checkpoint scored 0.934. Remaining architectural and schedule details are proprietary. Model weights will be released for noncommercial use (see **Code and model availability**).

### *4.3 Few-shot calibration*

Few-shot calibration fits a calibration head on a frozen encoder using registered tile pairs from a small calibration set. Here the set comprised 25 specimens with Leica reference scans: 15 University of Chicago specimens (five each of head-and-neck, colorectal and breast cancer; AT2, Grundium and Olympus comparison scans) and the ten PAR tempering specimens (Grundium and Hamamatsu comparison scans). For each calibration source and comparison scanner, a transform mapping comparison to reference embeddings was first estimated by regression regularized toward the identity; these transforms served only as training targets. A calibration head of the same content-conditioned form as the correction head was then trained to reproduce the transformed embeddings from the comparison embeddings while leaving reference embeddings unchanged, so that no scanner identity is needed at inference. For stock H-optimus-0 the head was initialized as the identity; for Mettle-1.5 it replaced the tempering correction head and was initialized to reproduce it. Training length was selected by five-fold specimen-held-out cross-fitting within the calibration set, and the final head was refitted on all 25 specimens. No outcome labels, downstream predictions or evaluation specimens were used.

### *4.4 Image acquisition*

The University of Chicago cross-scanner cohort was scanned on a Leica GT450 reference and rescanned on Aperio AT2, Grundium Ocus 20, and Olympus VS200 devices. The Leica reference images had been converted to SVS with JPEG compression and a regenerated image pyramid; this conversion is part of the reference acquisition.

### 4.5 Whole-slide image processing

Whole-slide images were processed with Slideflow[44], using libvips for slide reading. Tiles of 224 × 224 pixels were extracted without stain normalization and stored in lossless PNG format, except SICAPv2 tiles, which were stored as JPEG (quality 95). Tiles spanned 224 μm (nominal 10×) for HPV prediction, the confounding experiment, image perturbations, colorectal MSI and prostate grading, and 112 μm (nominal 20×) for the breast signatures; the calibration set and the unseen-scanner panel used both magnifications. The same tiles were used for every arm. Every University of Chicago rescan was registered at the pixel level to the specimen's reference scan. At least five corresponding tissue landmarks were placed manually on each rescan–reference pair, each correspondence was refined by automated translation alignment of the surrounding image windows, and an affine transform, with scale set by the ratio of pixel sizes, was fitted to the landmark pairs by least squares and used to map the reference tile grid onto the rescan. Tissue quality control was performed on the reference scan and transferred to each rescan through the registration, so that all acquisitions of a specimen were tiled at identical tissue locations with identical tile selection. Differences between acquisitions therefore do not arise from shifts in the tile grid or from differing quality control.

### 4.6 HPV prediction and paired-scanner evaluation

HPV status was defined by polymerase chain reaction for the viral genes E6 and E7 or by p16 immunohistochemistry. When both were available, the PCR result determined status. Predictors were trained on 345 TCGA head-and-neck slides[33], with 114 site-held-out validation slides (113 evaluable), which were not used for model selection.

For each arm, ten gated instance-cluster MIL models (CLAM)[14,15] were trained with identical settings and the same ten pinned seeds: 20 epochs, 512-tile bags, batch size 64, dropout 0.25, bag-loss weight 0.7, weight decay $10^{-5}$ and class-weighted loss. Probabilities were averaged across the ten models. A historical stock series with byte-identical hyperparameters and unrecorded random seeds gave a discrepancy of 0.106 Mascaret models used a validation-selected configuration (bag-loss weight 0.85, eight seeds).

The paired cohort comprised 101 enrolled cases; five used for calibration were excluded, and 88 had the Leica reference. Twenty-five rescans with focus-capture failures (whole-slide blur flagged by routine scanning quality control) were excluded. Scanner discrepancy was computed separately for each rescanning device as the root-mean-square difference between within-specimen rescan and reference probabilities; the three device values were averaged with equal weight. Supplementary analyses used logit-scale discrepancy (probabilities clipped to $[10^{-6}, 1 - 10^{-6}]$) and discrepancy divided by each system's reference-score standard deviation. Among the 63 specimens with all four acquisitions, they also used the two-way random-effects, absolute-agreement, single-measure intraclass correlation. Classification analyses used a conventional threshold of 0.5. Supplementary analyses also used model-specific Youden thresholds derived retrospectively in the Chicago outcome cohort (362 cases).

### 4.7 Controlled scanner–disease confounding for HPV prediction

Within each AT2-scanner pair, fixed pools of 75 (Grundium), 88 (Leica) and 89 (Olympus) cases with both acquisitions were used; calibration cases were excluded. Mascaret features (the developers' recommended normalized class token; base encoder pretrained on data including TCGA) were run through the identical design, allocations, seeds and training settings. For each case, one acquisition entered training, chosen to impose a nominal scanner–HPV association of 0, 0.25, 0.5, 0.75 or 1 (with realized $|V|$ within 0.01 of nominal). Ten allocations per level were generated, five in each association direction, with three fits per allocation, identical across arms. As a post hoc readout, the share of TCGA patients called HPV-positive at 0.5 was compared between association directions with the same hierarchical bootstrap.

Predictors used gated instance-cluster MIL with 512 latent and 256 attention features, 20 epochs, 512-tile bags, batch size 64, dropout 0.25, class-weighted loss, instance-level support-vector-machine loss (k = 8), bag-loss weight 0.7, weight decay $10^{-5}$ and a one-cycle schedule, without a validation cohort. Each model was evaluated on 458 TCGA slides from 436 patients (50 HPV-positive), with slide probabilities averaged within patients. Readouts, reported together, were mean AUROC at maximal association (compared between arms on identical allocations and fits), the loss in mean AUROC from zero to maximal association and the difference between directions in the proportion of TCGA patients called HPV-positive.

### *4.8 Prostate grading*

Grade-group models were trained on 568 slides from TCGA-PRAD[34] and SICAPv2[35] using three fixed folds (381/187, 378/190 and 377/191 training/validation slides). Attention-based MIL models[14] were trained for 30 epochs with learning rate $10^{-4}$, weight decay $10^{-5}$, 512-tile bags, batch size 32, one-cycle scheduling and weighted cross-entropy, selecting checkpoints by quadratic weighted kappa. Each of 40 fits per arm averaged threefold models, and predicted grade was the most probable class.

The evaluation used PAR prostate biopsies[32] acquired on Grundium Ocus 40 and Hamamatsu NanoZoomer 2.0 HT scanners. PAR provides independent, non-adjudicated ISUP grades from up to three pathologists[32]: pathologist 1 graded every slide, pathologist 2 all but two, and pathologist 3, a uropathologist, a grade-stratified subset. The reference was pathologist 1's ISUP grade group (1–5). Of 174 available specimens, nine used in tempering and calibration (the tenth was not among the 174) and six sharing a case prefix with them were excluded, leaving 159 specimens in 100 case families (specimens sharing a case identifier). A sensitivity analysis excluding only the nine gave 165 specimens with consistent results. Cross-scanner agreement used quadratic weighted kappa, exact agreement and agreement within one grade. Reference agreement was calculated per scanner and averaged. Agreement of each arm and of the tempered control with pathologists 2 and 3, and between pathologists, was computed on the specimens each graded (157 and 39) on the ISUP scale including benign (0), which the models could not assign; calls at grade group 5 and the control's agreement with pathologists 2 and 3 were added post hoc.

### *4.9 Colorectal MSI and breast signatures*

MSI models were trained on TCGA colon and rectal adenocarcinoma slides[36]. MSI status was taken from polymerase chain reaction testing where available and otherwise from concordant sequencing-based calls (MANTIS[45] and MSIsensor[46]). Slides were grouped by tissue source site into five folds; four (473 slides from 468 patients at 25 sites; 65 MSI-high) were used for training and the fifth (101 slides from 98 patients at 11 sites; 12 MSI-high), fixed before training, for site-held-out validation (reporting only). For each arm, ten models were trained with the settings and seeds used for HPV, and probabilities were averaged. For the University of Chicago colorectal cohort, MSI status was determined by the institutional next generation sequencing assay (OncoPlus). Of 86 selected specimens, 78 were scanned (16 MSI-high), and the five calibration specimens (two MSI-high) were excluded, leaving 73. Whole-slide features were scored as for HPV. Eleven of the 73 lacked the Leica reference scan (one also lacked the Grundium rescan), so discrepancy used the 62 specimens (13 MSI-high) with the reference and all three rescans (186 pairs), and AUROC was averaged over the four acquisitions.

The breast models were two continuous outputs of a previously described model predicting gene-expression signatures from H&E histology[5]. One is the bronchioid signature[37], identified as predictive of chemotherapy benefit in early breast cancer[5]; the other approximates the 21-gene recurrence score[38]. For each specimen, 100 tissue locations were selected on the reference acquisition by stratified attention of the stock models, with attention percentiles averaged across 25 models. Each scan's 100 locations were scored as one bag by each model. The same registered locations were used for all acquisitions and arms. After excluding the five calibration specimens, 77 breast specimens remained. Because these models were trained on stock features and applied unchanged, the main measure was discrepancy divided by each arm's reference-score standard deviation.

### *4.10 Unseen-scanner panel*

Six head-and-neck specimens, four of which are also in the paired HPV cohort and one of which is a calibration specimen (excluded in a sensitivity analysis), were scanned on a Leica Aperio GT450 reference and six comparison devices: Leica Aperio AT2, Olympus VS200, and Grundium Ocus 20 devices of the models used for calibration, and an Aperio XT, a 3DHISTECH Pannoramic Scan 150 and a Philips IntelliSite UFS, which were not. Dense 10× tiles were scored with each arm's ten refitted HPV models (as in Fig. 2), and discrepancy relative to the reference was averaged within device groups. With six specimens, no intervals were computed.

### *4.11 Image perturbations*

Fifty TCGA validation slides from 46 participants (163,073 tiles) were selected across the stock prediction range. Each system's ten HPV models were applied to an unperturbed control and to seven perturbations: JPEG re-encoding at quality 90 and 80, two Macenko stain targets[23] and three fixed color lookup tables, following synthetic stress-test designs[47]. For calibrated Stock, the calibration head was applied to the perturbed stock features. Drift was the root-mean-square change in ensemble probability from each system's own control.

### *4.12 Public benchmarks*

PathoROB, CRoMa, HEST and EVA were run with their official implementations and default settings, without additional tuning. PathoROB robustness indices[21] used class-token features across Camelyon, TCGA and Tolkach cohorts; published leaderboard values for other models, including Mascaret[27], use concatenated class-token and mean patch-token features, in which Mettle-1.5 was also scored. CRoMa[39] used concatenated class-token and mean patch-token features; Mettle-1.5 was scored with the released CRoMa code and ranked against the 26-model public leaderboard. HEST[40] used nine tasks with principal-component (256) ridge regression. BACH[41] used five fits with identical 268/132 image splits, and EVA tasks used the standard EVA framework[42]. Stock HEST values are published class-token references.

### *4.13 Statistical analysis*

Confidence intervals were 95% percentile intervals from 2,000 paired bootstrap resamples, drawn at the level of:

- specimens (paired-scanner analyses);
- cases (Chicago outcome cohort);
- case families (prostate);
- slides (perturbations; 10,000 resamples).

Confounding intervals used a paired hierarchical bootstrap over TCGA patients, allocations within direction and fits, shared across encoders. Proportions used Wilson intervals. Each application was summarized by one main measure: scanner discrepancy averaged over devices (macro) for refitted HPV and MSI models, AUROC at maximal association and the between-direction difference in calls for confounding, cross-scanner weighted kappa for prostate, standard-deviation-scaled discrepancy for the breast signatures. Two-sided P values were obtained by inverting the paired percentile bootstrap ($P = 2 \cdot \min(\Pr[\Delta \leq 0], \Pr[\Delta \geq 0])$ over the same resamples used for the confidence intervals). Analyses were conducted after model locking; no adjustment for multiplicity. Comparisons were descriptive and reported with 95% CIs and unadjusted P values.

## Code and model availability

Mettle-1.5 weights are released at https://huggingface.co/slideflow-labs/mettle-1.5 under CC BY-NC-ND 4.0. Tempering and calibration training code and calibration heads are proprietary and not released.

## Figures

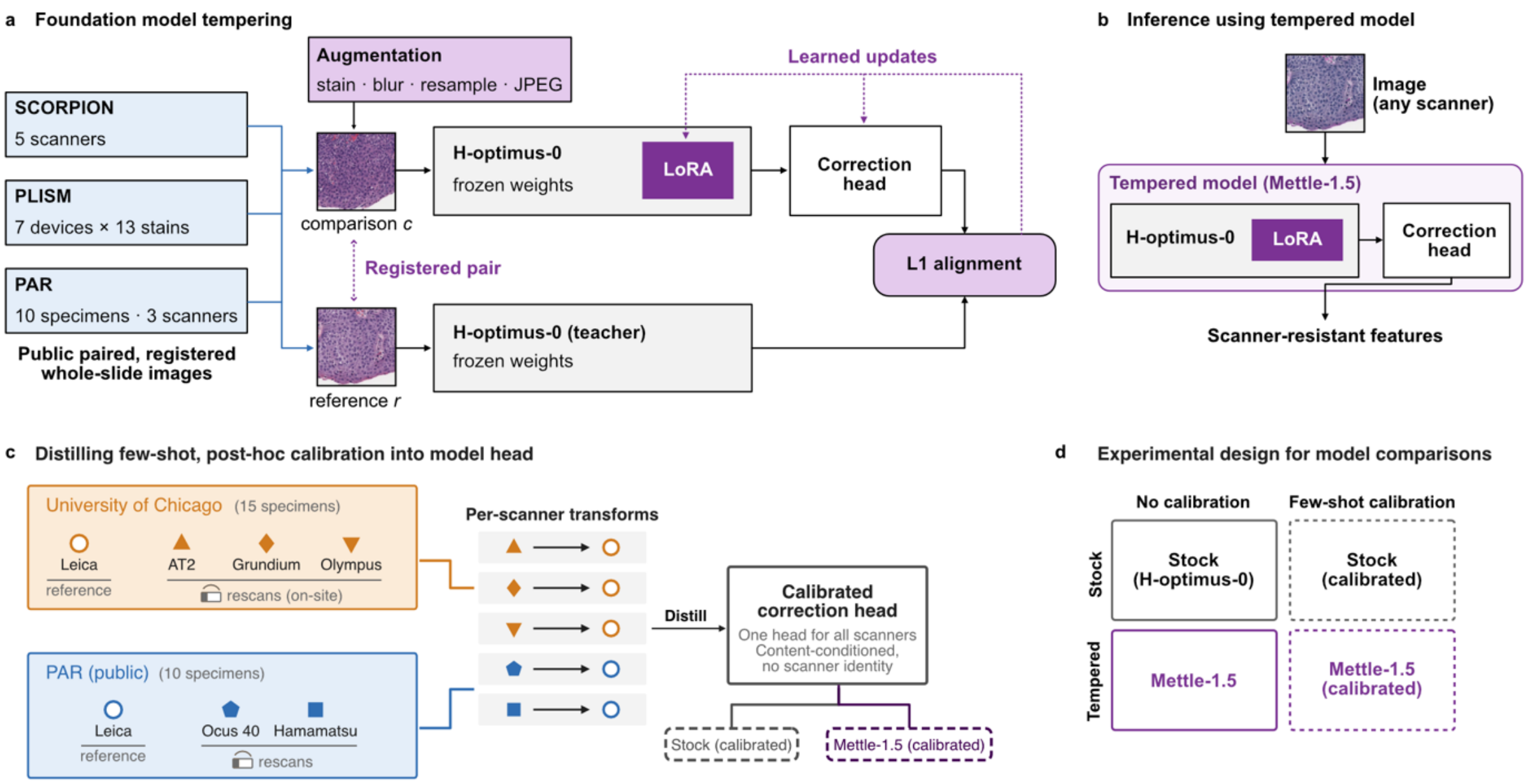

**Fig. 1 | Schematic of tempering and calibration method and datasets. (a)** Tempering protocol. Public paired, registered whole-slide images supply a comparison view (*c*) and a reference view (*r*) of the same tissue location; *c* is further augmented (stain, blur, resampling, JPEG). Low-rank adapters (LoRA) on frozen H-optimus-0 weights, followed by a correction head, embed *c*, and the original model (teacher) embeds *r*; training minimizes the L1 distance between the embeddings and updates only the adapters and the head (dotted purple arrows). **(b)** During inference, Mettle-1.5, with adapters and correction head merged into one model, embeds a single image without scanner metadata. **(c)** Few-shot calibration. Fifteen University of Chicago specimens (orange) and ten public PAR specimens (blue; those used in tempering, a) were scanned on a Leica reference scanner and rescanned on comparator scanners with pixel-level image registration. Per-scanner feature space transforms from rescan to reference embeddings were distilled, as training targets, into a content-conditioned calibration head that needs no scanner identity at inference. **(d)** The four arms compared throughout this study: stock H-optimus-0 or tempered Mettle-1.5, each with or without the calibration head; for Mettle-1.5, the calibration head replaces the tempering correction head and is initialized from it. Calibrated arms (dashed) require registered slides from the scanners to be corrected.

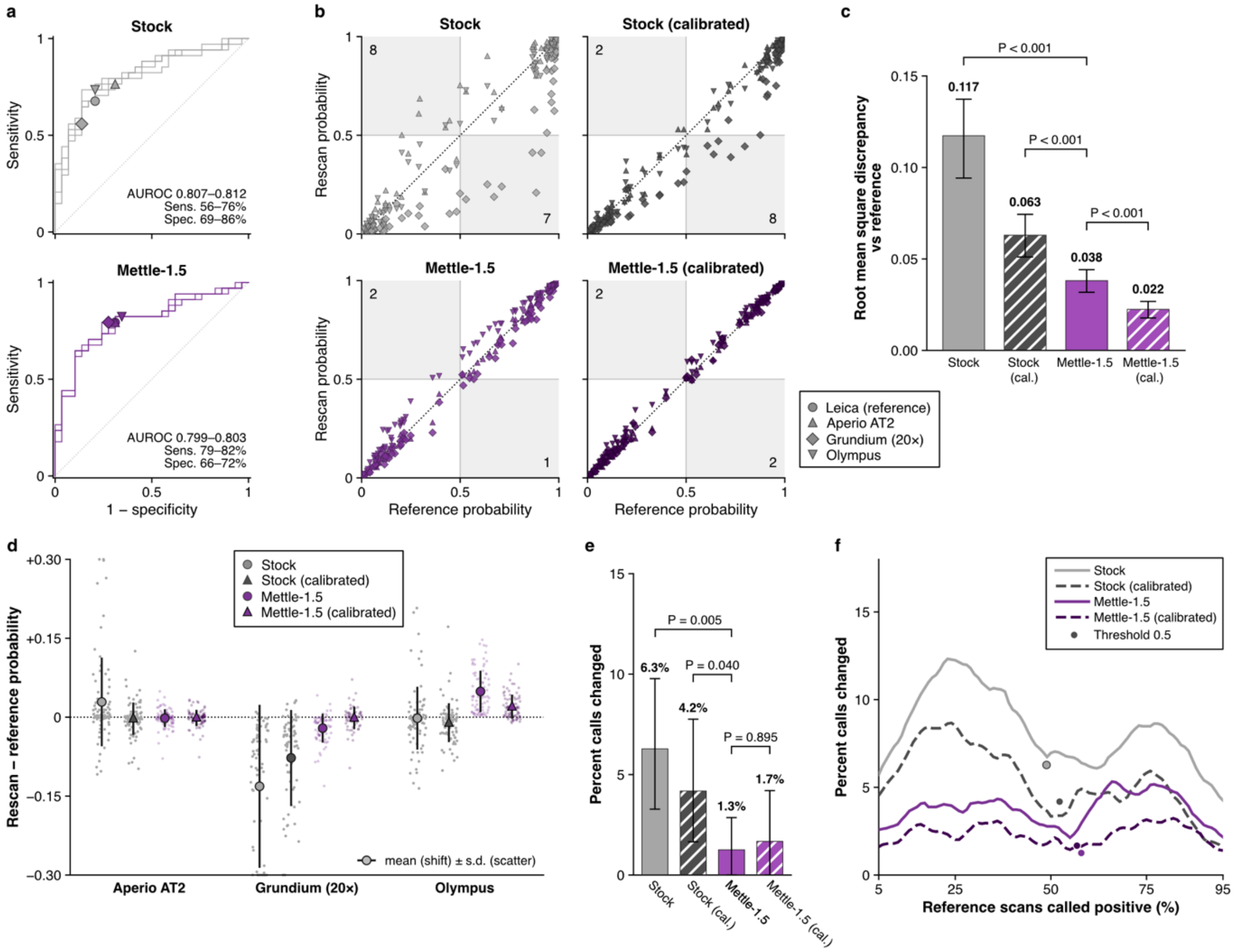

**Fig. 2 | Tempering increases cross-scanner consistency of HPV status prediction in head and neck cancer. (a)** ROC curves for Stock and Mettle-1.5 on the 63 University of Chicago test specimens (34 HPV-positive) scanned with four different scanners. Markers indicate operating points for the same 63 cases at a fixed 0.5 threshold with each distinct scanner. Sensitivity and specificity across scanners is shown as a range. Sensitivity and specificity significantly differed across scanner acquisitions for models trained with Stock ($P = 0.002$ and 0.017), but not with models using Mettle-1.5 ($P = 0.39$ and 0.26). **(b)** Ensemble HPV probability on each rescan versus the Leica reference (239 pairs, 88 specimens). **(c)** Root mean square (RMS) discrepancy in prediction probability between reference and comparator scanners, averaged over the three comparator scanners. Mettle-1.5 minus Stock is a 68% reduction (59–74%, $P < 0.001$). **(d)** Rescan minus reference probability for every pair, by scanner: dot, mean (systematic shift); bar, ±1 s.d. (scatter). **(e)** Calls changed at fixed 0.5 threshold over all 239 pairs. Mettle-1.5 decreases the percent of changed calls relative to Stock (absolute reduction 5%, $P = 0.005$). **(f)** Calls changed when each arm's threshold calls the same percentage of reference scans positive (5–95%): lines, centered moving average over 11 levels (±5 percentage points); dots, the 0.5 threshold. Each arm averages ten models trained with identical seeds.

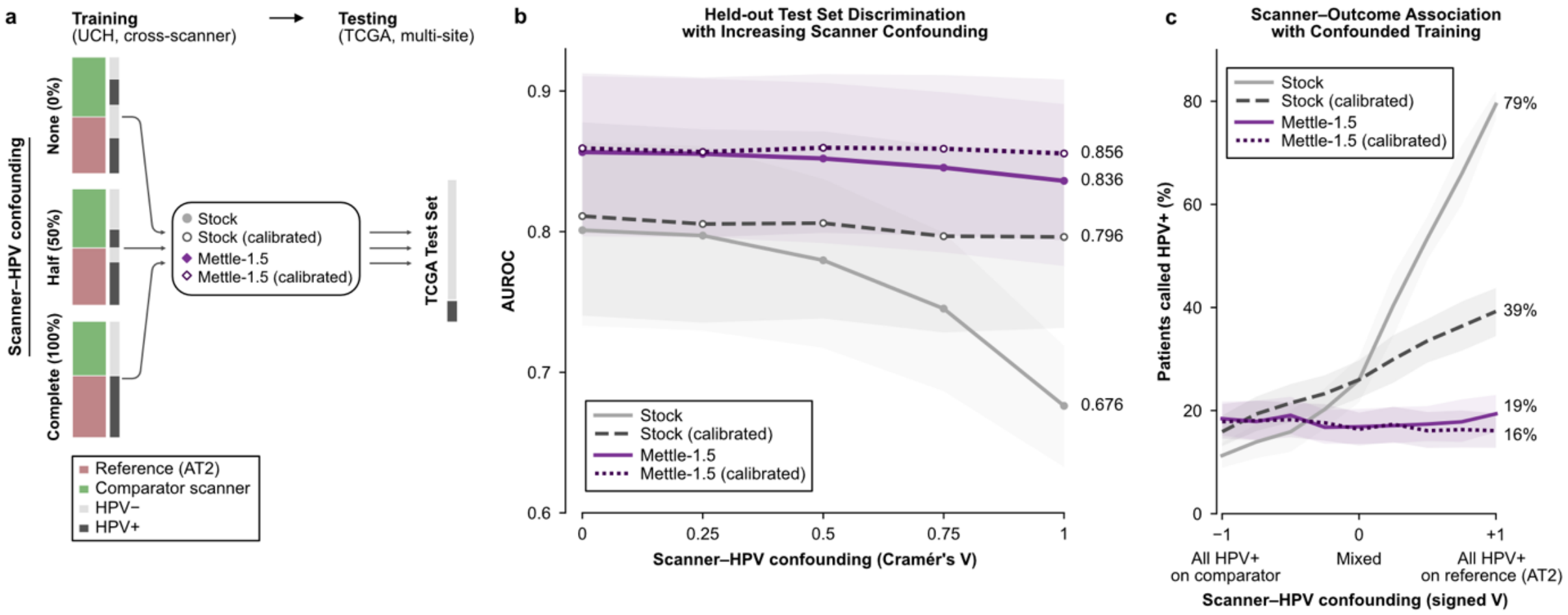

**Fig. 3 | Models trained on tempered features better preserve performance under scanner-outcome confounding in the training data.** **(a)** HPV prediction models were trained on University of Chicago head and neck cancer cases scanned on an AT2 and a comparator scanner. The acquisition used for each case was chosen to construct training set with a scanner–HPV association (Cramér's V) from 0 to 1. (**b**) Mean test-set TCGA AUROC after training with variable confounding (V from 0 to 1). Shaded areas indicate 95% CI. (**c**) Percent of TCGA patients called HPV-positive (probability ≥ 0.5) after scanner-confounded training (positive V indicating HPV-positive training images from the AT2; negative, from the comparator scanner).

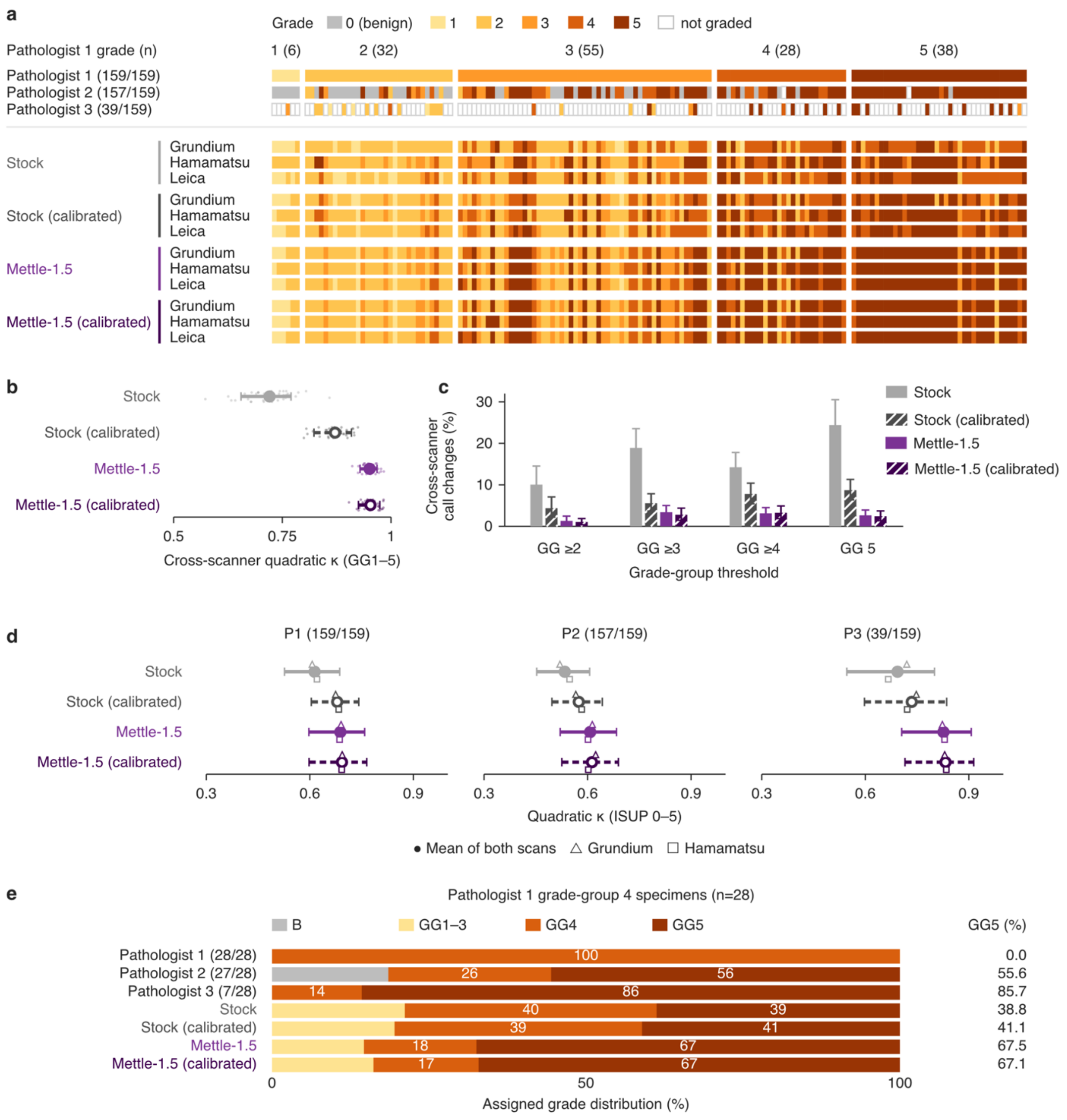

**Fig. 4 | Tempering makes prostate grade groups reproducible across scanners and increases agreement with pathologists.** 159 PAR biopsy specimens (100 case families) scanned on a Grundium Ocus 40 and a Hamamatsu NanoZoomer 2.0 HT; 40 grading models (three-fold ensembles) per arm. (a) Tiles show individual specimens. Top row shows each pathologist's assigned grade (grey, benign; white, not graded). Bottom rows show model predicted grade from each encoder arm for each specimen. (b) Cross-scanner quadratic weighted κ between predicted grades for paired Ocus 40 and NanoZoomer scans. Small points show individual models, large points show means over models. (c) Percentage of specimens whose call at a grade-group threshold (≥2, clinically significant cancer; ≥3, unfavorable intermediate risk; ≥4, high

risk; 5) differed between scanners. (d) Quadratic weighted κ between model-assigned grades and each pathologist's grades, (P1 159 specimens, P2 157, P3 39). Triangles, Ocus 40 scans; squares, NanoZoomer scans; circles, mean of the two. (e) Grades assigned to pathologist 1's 28 grade group 4 specimens by pathologists 2 and 3 and by each arm (both scans, all models); right, percentage graded 5. Reference, pathologist 1's ISUP grade group. Error bars show 95% bootstrap CIs over case families and models.

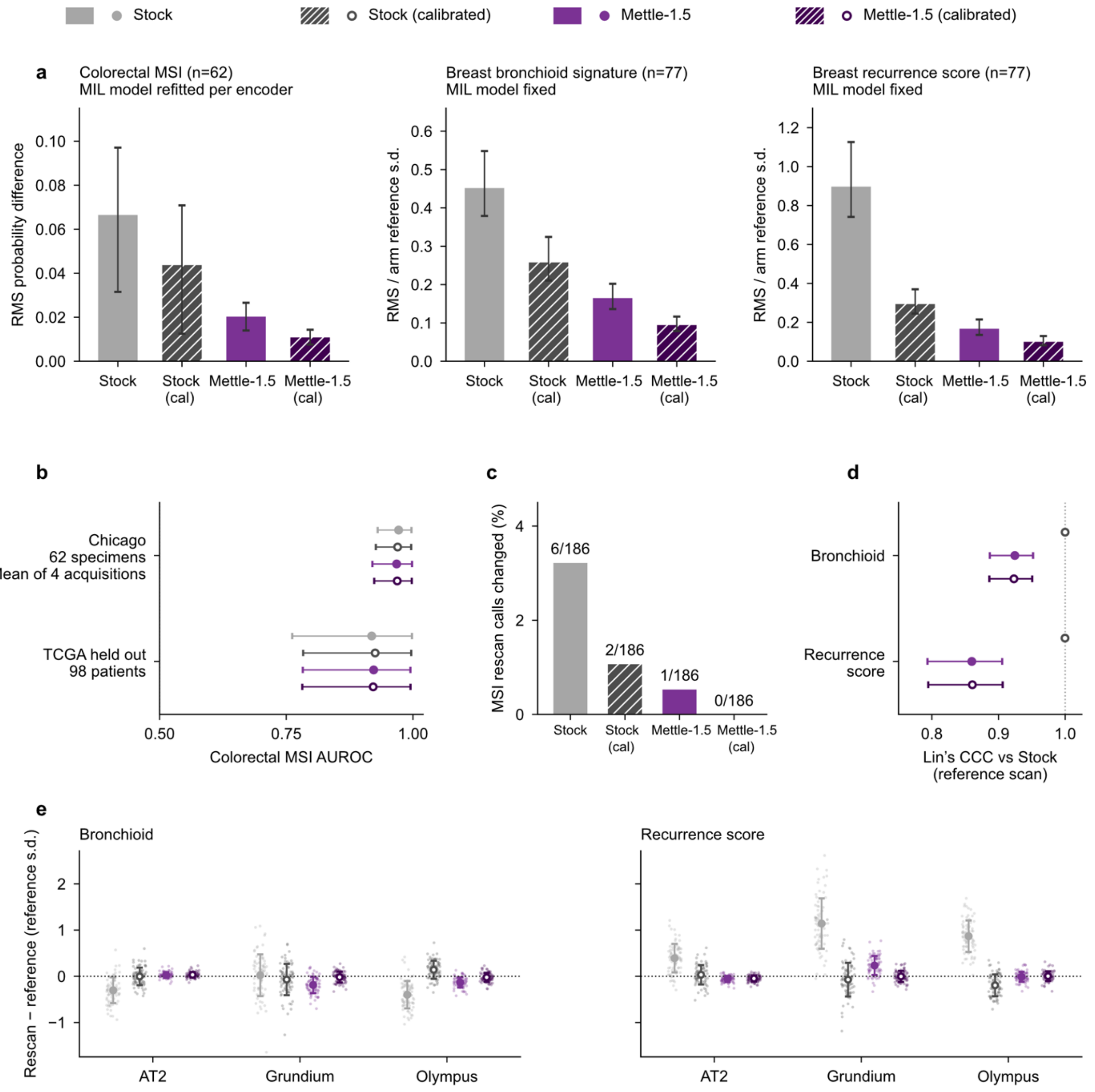

**Fig. 5 | Tempering improves cross-scanner consistency of colorectal MSI prediction and of frozen breast gene-expression models.** (a) RMS discrepancy between rescan and reference predictions. Left, MSI probability on 62 University of Chicago specimens (186 pairs), models trained on each arm's features. Middle and right, bronchioid signature and recurrence-score approximation on 77 breast specimens at 100 registered locations per acquisition using frozen published models trained on Stock features and applied to each arm's features unchanged, with discrepancy divided by each arm's reference-scan s.d. (b) MSI AUROC on the Chicago cohort (mean of four acquisitions; 13 MSI-high) and 98 held-out TCGA patients. (c) MSI calls changed at the 0.5 threshold over 186 pairs. (d) Lin's concordance between each arm's breast scores and Stock scores on the reference scan. (e) Rescan minus reference score for every pair, by scanner: dot, mean; bar, ±1 s.d. Error bars show 95% bootstrap CIs over paired specimens and seeds.